\documentclass[twocolumn,showpacs,preprintnumbers,amsmath,amssymb]{revtex4}

\usepackage{graphicx}
\usepackage{dcolumn}
\usepackage{bm}

\begin{document}
\title{Spintronic properties  of  one-dimensional electron gas in graphene armchair ribbons }
\author{J. W. Lee, S. C. Kim,  and S. -R. Eric Yang\footnote{corresponding author, eyang812@gmail.com}}
\affiliation{
Physics Department, Korea  University, Seoul 136-701, Korea\\
}

\begin{abstract}
We have investigated, using effective mass approach (EMA), magnetic
properties of a one-dimensional electron gas in graphene armchair
ribbons when the electrons of occupy only the lowest conduction
subband. We find that magnetic properties of the one-dimensional
electron gas may depend sensitively on the width of the ribbon. For
ribbon widths $L_x=3Ma_0$, a critical point separates ferromagnetic
and paramagnetic states while  for $L_x=(3M+1)a_0$ paramagnetic
state is stable ($M$ is an integer and $a_{0}$ is the length of the
unit cell). These width-dependent properties are a consequence of
eigenstates that have  a subtle width-dependent mixture of
$\mathbf{K}$ and $\mathbf{K'}$ states, and can be understood by
examining  the wavefunction overlap that appears in the expression
for the many-body exchange self-energy. Ferromagnetic and
paramagnetic   states may be used for  spintronic purposes.

\end{abstract}
\maketitle

\section{Introduction}

Generation of spin-polarized current is of significant importance
both scientifically and technologically. A spin-polarized current
would emerge naturally from  ferromagnetic materials.  In order to
achieve a spintronic device, it is important to find non-magnetic
materials where a spin-polarized current can be  flowed without
becoming depolarized.   It has become possible to induce and detect
spin polarization in  non-magnetic
semiconductors\cite{NatP,Nat,Yang}. In this paper we explore the
possibility that  graphene\cite{Geim} field effect
transistors\cite{Sch} based on armchair nanoribbons can  be used to
generate spin-polarized currents. The conduction electrons in
carbon-based materials can move very long distances without
scattering due to their small spin-orbit coupling and low hyperfine
interaction.   Recently magnetic effects using graphene  edges
states of an armchair ribbon have been explored\cite{Lin}. In our
work we investigate bulk magnetic properties of a one-dimensional
electron gas of armchair ribbons, where electrons interact via
long-range Coulomb interaction. It is possible that a ferromagnetic
state may be stable in such a system as the investigation of
one-dimensional Hubbard model with nearest neighbor
electron-electron interaction suggests\cite{Dau}.

\begin{figure}[!hbpt]
\begin{center}
\includegraphics[width=0.3\textwidth]{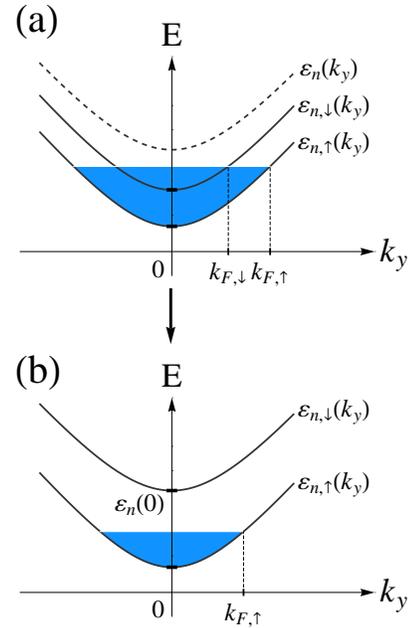}
\caption{(a) Electron density is such that  the exchange self-energy
is smaller than the Fermi energy  and the electron gas is partially
spin-polarized.  We assume that, among conduction subbands,  only
the lowest energy conduction subband is occupied with electrons.
Dashed line indicates spin degenerate subband energy in the absence
of electron-electron interactions. Spin-up and -down subbands are
shown. (b) For a smaller electron density the exchange self-energy
can be  larger than the Fermi energy and the electron gas is fully
spin-polarized.} \label{fig:splitting}
\end{center}
\end{figure}

Graphene armchair ribbons\cite{Son,Brey1,Neto} have several special
properties that are well suited for spintronic applications.
According local density approximation (LDA) when the width of an
armchair ribbon is $L_x=3(M+1)a_0$ or $L_x=3Ma_0$ a gap exists in
the energy spectrum \cite{Son} (a rather small gap exists when
$L_x=(3M+2)a_0$, and this case will not be considered here). The
other property is that boundary condition on the armchair edges
admix $\mathbf{K}$ and $\mathbf{K'}$ valleys, and eigenstates are
mixture of $\mathbf{K}$ and $\mathbf{K'}$ states forming
one-dimensional subbands\cite{Brey1} (this is in contrast to
parabolic and cylindrical potentials\cite{Park1,Park2,Kim}, where
the mixing is rather small). When the system is doped electrons
occupy these subbands and a one-dimensional electron gas forms, see
Fig.\ref{fig:splitting}. The other unique property is a rather small
value of the dielectric constant ($\epsilon\sim 1$), which makes
effects of Coulomb interaction effects strong. Many-body
self-energies are thus one order of magnitude larger in comparison
to  the corresponding values of an electron gas of ordinary
semiconductors with $\epsilon\sim 10$.

\begin{figure}[!hbpt]
\begin{center}
\includegraphics[width=0.35\textwidth]{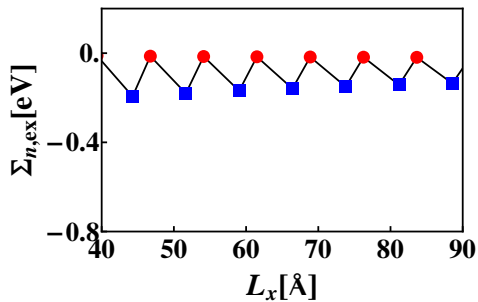}
\caption{ Exchange self energies of doped graphene for ribbon widths
$L_x=3(M+1)a_0$ (circles) and $L_x=3Ma_0$ (squares), where
$a_0=2.46\AA$ and $M$ is an integer. The effective mass approach is
used.  Here the dielectric constant is $\epsilon=3$ and the
dimensionless Fermi wavevector is $k_F a_0=0.07$.} \label{ExchSE}
\end{center}
\end{figure}

\begin{figure}[!hbpt]
\begin{center}
\includegraphics[width=0.35\textwidth]{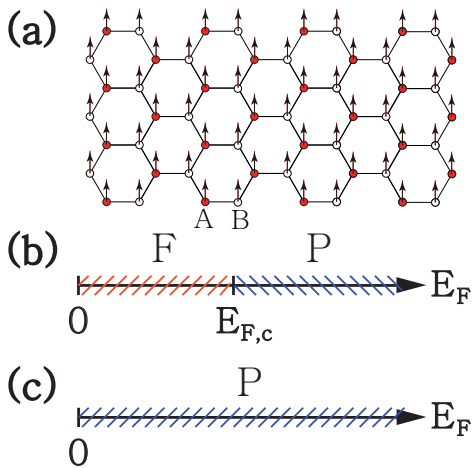}
\caption{ (a) Bulk ferromagnetic  state of one-dimensional electron
gas in a graphene armchair ribbon.  We assume that only  the
conduction subband of a graphene armchair ribbon that is closest in
energy to $E=0$ is occupied. The average spin value per carbon site
is shown as vertical arrows (Only spins of electrons belonging to
the lowest conduction subband are shown).  (b) Phase diagram as a
function of the Fermi energy $E_F$ for the length of ribbon
$L_x=3Ma_0$.  $E_F$ is measured from the bottom of the conduction
subband.  F (P) stands for ferromagnetic (paramagnetic) state. (c)
Phase diagram for $L_x=(3M+1)a_0$.} \label{phase}
\end{center}
\end{figure}

In this work we assume that a gap separates conduction and valence
subbands, and that, among conduction subbands,  only the lowest
energy conduction subband is occupied with electrons.  We calculate
the many-body exchange self-energy $\Sigma_{ex}$ of a spin-polarized
one-dimensional electron gas of such a system for relatively large
values of the width $L_x\ge 45\AA$, where the LDA and EMA results
agree approximately (see Sec.II). The width dependence of
$\Sigma_{ex}$ is shown in Fig.\ref{ExchSE}. Using these results we
find that, for ribbon widths $L_x=3Ma_0$, a critical point separates
ferromagnetic and paramagnetic states while paramagnetic state is
stable for $L_x=(3M+1)a_0$.  These results are illustrated in
Fig.\ref{phase}. This dependence on the width can be understood by
examining  the wavefunction overlap that appears in the expression
for the exchange self energy.  The effect is a consequence of
eigenstates that have a subtle width-dependent mixture of
$\mathbf{K}$ and $\mathbf{K'}$ states. The large difference in the
values of dielectric constants of graphene and ordinary
semiconductors cannot explain these width-dependent magnetic
properties.

\section{Effective mass approximation and model Hamiltonian}

\begin{figure}[!hbpt]
\begin{center}
\includegraphics[width=0.35\textwidth]{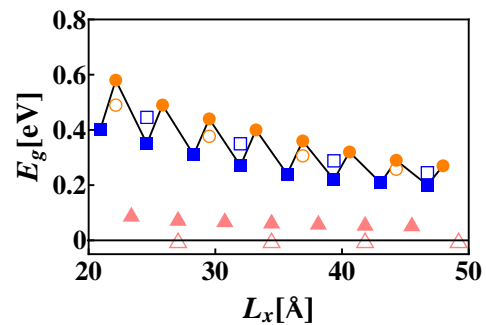}
\caption{Size of gap $E_g$ of undoped graphene is shown. LDA
results\cite{Son} for widths $L_x=3(M+1)a_0$ (filled circles),
$L_x=3Ma_0$ (filled squares), and $L_x=3(M+2)a_0$ (filled
triangles).  EMA results for the same widths are open squares,
circles, and triangles.} \label{DFT}
\end{center}
\end{figure}

Effective mass approach can describe numerous physical properties of
graphene approximately. It can be derived from tight binding
method\cite{Ando}. Fig.\ref{DFT} displays the value of the gap of
armchair ribbons as a function of the width. Both LDA\cite{Son} and
EMA results are shown for {\it undoped} armchair ribbons. We see
that the LDA values of the magnitude of the gap are in rough
agreement with those of the EMA results, and the agreement becomes
better for larger values of $L_x$. However, there are several
deviations. The EMA value of the gap is zero for $L_x=(3M+2)a_0$
while that of LDA is small but non-zero. Also as $L_x$ changes
between $3(M+1)a_0$ and $3Ma_0$ the LDA value of the gap displays
small oscillations while the EMA results do not. LDA  includes
electron-electron interaction effects and goes beyond nearest
neighbor hopping while EMA  does not. However, when $L_x$ is larger
than $45\AA$ and $L_x=3(M+1)a_0$ or $L_x=3Ma_0$ the EMA results are
approximately correct. We will employ EMA to compute the approximate
exchange self energy under these conditions.

\begin{figure}[!hbpt]
\begin{center}
\includegraphics[width=0.4\textwidth]{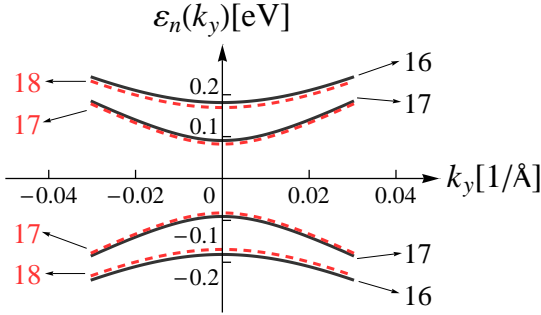}
\caption{Conduction and valence subbands with energies closest near
$E=0$.  Solid lines are for $L_x=3Ma_0=24a_0$. Dashed lines are for
$L_x=(3M+1)a_0=25a_0$.  The numbers beside the curves indicate
values of subband index $n$.} \label{subband}
\end{center}
\end{figure}

We choose $\mathbf{K}$ and $\mathbf{K'}$ valleys as
$\mathbf{K}=\frac{2\pi}{a_0}(\frac {1}{3},\frac{1}{\sqrt3})$ and
$\mathbf{K'}=\frac{2\pi}{a_0}(-\frac {1}{3},\frac{1}{\sqrt3})$. The
ribbon is along y-axis. The envelope wavefunctions
$[\psi_A,\psi_B,]$ and $[\psi_A^{'},\psi_B^{'}]$ can be combined
into $\Psi=(\psi_A,\psi_B,-\psi_A^{'},-\psi_B^{'})$, which
satisfies the Hamiltonian
\begin{eqnarray}
H_0= \gamma a_0 \left(\begin{array}{cccc}
0 & \mathbf{k}_x - i\mathbf{k}_y & 0 & 0\\
\mathbf{k}_x + i\mathbf{k}_y & 0 & 0 & 0\\
0 & 0 & 0 & -\mathbf{k}_x - i\mathbf{k}_y\\
0 & 0 & -\mathbf{k}_x + i\mathbf{k}_y & 0\\
\end{array}\right).
\nonumber\\
\end{eqnarray}
Wavevector $k_{x,y}$ is measured from  $\mathbf{K}$ and
$\mathbf{K'}$ in the upper and lower Hamiltonians, respectively. The
hard wall boundary conditions\cite{Tang} on each of the total A and
B components of the wavefunction at the armchair edges $x=0$ and
$x=L'_x=L_x+a_0$ admix $\mathbf{K}$ and $\mathbf{K'}$ valleys. Note
that envelope wavefunctions along the x-axis are $\textrm{e}^{\pm
ik_nx}$ and have the opposite wavevectors $k_n$ and $-k_n$ for
$\mathbf{K}$ and $\mathbf{K'}$.  The wavefunction $\Psi$ of
conduction subbands are\cite{Brey1}
\begin{eqnarray}
\Psi_{n}(x,y,k_y)= \frac{\textrm{e}^{ik_yy}\theta(x)}{2\sqrt{L'_x}
\sqrt{L_y}} \left(\begin{array}{cccc}
-\textrm{e}^{-i\theta_{k_n,k_y}}\textrm{e}^{ik_nx}\\
\textrm{e}^{ik_nx}\\
-\textrm{e}^{-i\theta_{k_n,k_y}}\textrm{e}^{-ik_nx}\\
\textrm{e}^{-ik_nx}\\
\end{array}\right),
\label{wave}
\end{eqnarray}
where $\theta(x)= 1$ for $ 0\leq x\leq L'_x$ and $0$ otherwise. The
wavevector of the n'th subband is
$k_n=\frac{n\pi}{L'_x}-\frac{2\pi}{3a_0} $ and
$\theta_{k_n,k_y}=\textrm{Arctan}(k_y/k_n)$. The eigenenergy is
$\epsilon_{n}(k_y)=\gamma a_0\sqrt{k_y^2+k_n^2}$, where
$\gamma=\frac{\sqrt{3}}{2}t$ and $t=2.7$eV. Subband energies near
$E=0$ are plotted in Fig.\ref{subband} for two possibilities
$L_x=3Ma_0$ and $L_x=(3M+1)a_0$.  In this figure the value of $L_x$
is chosen so that in the first case $k_n>0$ while in the second case
$k_n<0$.

\section{Exchange self-energy}

We consider spin-polarized electrons in  the lowest energy
conduction subband with the electron density
$n_D=\frac{k_{F}}{\pi}$, where $k_{F} $ is the Fermi wavevector. In
our model, we ignore the inter-subband mixing by Coulomb
interaction, which is a standard approximation\cite{Haug}.  Then
Hartree-Fock {\it self-consistent} eigenfunctions are {\it plane
waves}\cite{Ash}, given by Eq.(\ref{wave}), and no corrections of
the kinetic energy are present. Note the Hartree self-energy will
cancel with the potential of uniform positive background.

\begin{figure}[!hbpt]
\begin{center}
\includegraphics[width=0.3\textwidth]{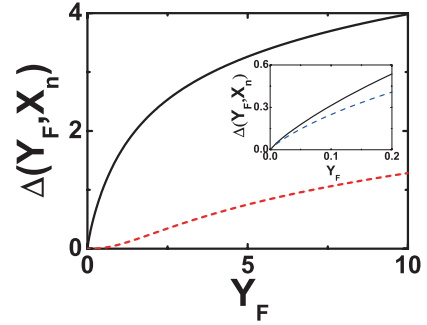}
\caption{Magnitude of dimensionless exchange self-energy
$\Delta(Y_F,X_n)$ as a function of $Y_F$ for $X_n=\pi/3$ (solid
line) and $-\pi/3$ (dashed line).  Inset: magnitude of the
dimensionless self-energy $\Delta(Y_F,X_n)$ as a function of $Y_F$
computed using Eq.(\ref{num}) (solid line) and Eq.(\ref{Meij})
(dashed line) for $X_n=\pi/3$.} \label{fig:exself}
\end{center}
\end{figure}

We denote  these wavefunctions  by $|k_y\rangle$. At $k_y=0$ the
exchange self-energy\cite{Ash} is
\newcommand{\ud}{\mathrm{d}}
\begin{eqnarray}\label{Coulomb interaction for graphene-1}
\Sigma_{n,ex}(0)&=&
-\frac{L_y}{2\pi}\int_{-k_F}^{k_F} \ud k_y \left\langle
0,k_y\left|\,\,\frac{e^2}{\epsilon|\mathbf{r_1}-\mathbf{r_2}|}\,\,\right|k_y,0\right\rangle,\nonumber\\
\end{eqnarray}
which can be written as $ \Sigma_{n,ex}(0)=-\frac{e^2}{ \epsilon
L'_x} \Delta(Y_F,X_n) $ with the dimensionless exchange self-energy
defined as
\begin{eqnarray}
&&\Delta(Y_F,X_n)\nonumber\\
&&=\textrm{\footnotesize{$\frac{1}{2\pi}\int_{-\infty}^{\infty} \ud
X \int_{-Y_F}^{Y_F}\ud Y \left(1+\frac{X_n}{\sqrt{X_n^2+Y^2}}\right)
\frac{(1-\cos X)}{\sqrt{X^2+Y^2}X^2}$}}.\nonumber\\
\label{num}
\end{eqnarray}
Here  $Y=k_yL'_x$,  $Y_F=k_{F}L'_x$ and $X_n=k_n L'_x$ are all
dimensionless.   In the case $L_x=3Ma_0$ we have $X_n=\pi
(n-2M-\frac{2}{3})$, and the conduction subband with the {\it lowest
energy} has the value $X_n=\pi/3 $ corresponding to the value
$n=2M+1$. For $L_x=(3M+1)a_0$ we have $X_n=\pi (n-2M-\frac{4}{3})$,
and the conduction subband with the {\it lowest energy} has the
value $X_n=-\pi/3 $ corresponding to $n=2M+1$.   In
Fig.\ref{fig:exself} $\Delta(Y_F,X_n)$ is plotted as a function of
$Y_F$ for  $X_n=\pi/3$ and $-\pi/3$. For $Y_F\ll 1$ we can
approximate
\begin{eqnarray}
\Delta(Y_F,X_n)
&\simeq& (1/\pi)\left(1+\frac{X_n}{\sqrt{X_n^2+Y_F^2}}\right)\left(\frac{-2}{Y_{F}^3}\right)\nonumber\\
&\times& \left(Y_{F}^2+2\,G_{1,3}^{2,1} \left(\begin{array}{c|c}
\begin{array}{cc}
5/2\\
1,\,2,\,3/2\\
\end{array}&\frac{Y_F^2}{4}
\end{array}\right)
\right)\nonumber\\
\label{Meij}
\end{eqnarray}
where $G$ is Meijer G-function\cite{Bat}. In Fig.\ref{fig:exself}
this approximate analytical result is compared with the exact
numerical result of Eq.(\ref{num}). As expected, the agreement
between the two are excellent for small values of $Y_F$.

Exchange self energies of doped graphene for ribbon widths
$L_x=3(M+1)a_0$  and $L_x=3Ma_0$ are shown in Fig.\ref{ExchSE}. Here
the dimensionless Fermi wavevector $k_F a_0=0.07$.  We notice that
the self energy displays  oscillations as $L_x$ changes between
$3(M+1)a_0$ and $3Ma_0$.  A similar effect was {\it also} observed
in the LDA result for the gap in the undoped case (compare
Figs.\ref{ExchSE} and \ref{DFT}).

Spontaneous spin splitting will occur in the lowest conduction
subband when $ |\Sigma_{n,ex}(0)|>E_{F}$, see
Fig.\ref{fig:splitting}. This condition is equivalent to
$\frac{e^2}{\epsilon \gamma a_0}\Delta(Y_F,X_n)>
\sqrt{X_n^2+Y_F^2}-|X_n| $.   At $X_n=\pi/3$    the inequality is
satisfied when $Y_F<11.309$ for $\epsilon=1$ and $Y_F<3.22$ for
$\epsilon=3$. The critical point between ferromagnetic and
paramagnetic states is shown schematically in Fig.\ref{phase}(b). At
$X_n=-\pi/3$ no critical point exists since $\Delta(Y_F,X_n)$ is too
small, see Fig.\ref{fig:exself}. In this case paramagnetic state is
stable, as illustrated in Fig.\ref{phase}(c).

Also we must require that the exchange self energy correction be
smaller than the gap $|\Sigma_{n,ex}(0)|<E_g=2\gamma a_0 |k_n|$.
This condition is equivalent to $ \frac{e^2}{\epsilon \gamma
a_0}\Delta(Y_F,X_n)< 2|X_n|=\frac{2\pi}{3}$.  We find at $X_n=\pi/3$
that this condition is satisfied when $Y_F<0.364$ for $\epsilon=1$
and $Y_F<2.439$ for $\epsilon=3$.

The dependence on the width can be understood by examining  the
wavefunction overlap that appears in the expression for the exchange
self energy. It is given by
$|\psi^*_n(x,y,0)\psi_n(x,y,k_y)|^2\propto 1+\cos\theta_{k_n,k_y}$
for $\mathbf{r_1}=\mathbf{r_2}$.   When $k_n<0$ the quantity
$\cos\theta_{k_n,k_y}=\frac{X_n}{\sqrt{X_n^2+Y^2}}<0$, which makes
the dimensionless self energy $\Delta(Y_F,X_n)$ smaller, see the
integrand in  Eq.(\ref{num}). This effect is a consequence of the
presence of the phase factors $\textrm{e}^{-i\theta_{k_n,k_y}}$ only
in certain components of the eigenstate wavefunctions (see
Eq.(\ref{wave})) and that  the value of $k_n$ for the lowest
conduction subband is {\it negative} when  $L_x=(3M+1)a_0$ (see
arguments below Eq.(\ref{num})). In other words it is a consequence
of a subtle width-dependent   mixture of $\mathbf{K}$ and
$\mathbf{K'}$ states in the  eigenstate wavefunctions.

\section{Summary and discussions}

As the comparison with the LDA result shows, when $L_x$ large and
for $L_x=3(M+1)a_0$ and $L_x=3Ma_0$ the result of EMA is
approximately correct. {\it Under} these conditions we have employed
EMA  to compute the approximate exchange self energy. We have  shown
that when {\it only} the lowest conduction subband of  a graphene
armchair ribbon is occupied magnetic properties of the
one-dimensional electron gas may depend sensitively on the width of
the ribbon. We find that, for ribbon widths $L_x=3Ma_0$, a critical
point separates ferromagnetic and paramagnetic states while
paramagnetic state is stable for $L_x=(3M+1)a_0$. This dependence is
in sharp contrast to one-dimensional electron gas of ordinary
semiconductors, and can be understood by examining  the wavefunction
overlap that appears in the expression for the exchange self energy
of an armchair ribbon. It reflects the fact that eigenstate
wavefunctions of an armchair ribbon  contain a non-trivial
width-dependent mixture of $\mathbf{K}$ and $\mathbf{K'}$ states.
The large difference in the value of dielectric constant between
graphene and ordinary semiconductors alone cannot explain our
width-dependent magnetic properties.  The magnitude of the exchange
spin splitting is of order $\frac{e^2}{\epsilon L'_x}$, and,
depending on the size of the Fermi wavevector $k_{F}$, it  can vary
in the range of $10-100$meV. As shown in Fig.\ref{fig:splitting} our
work suggests that  the degree of spin polarization  in a graphene
armchair ribbon may be controlled by changing   the gate voltage.

An estimation, using the usual mean field value, gives that Curie
temperature $k_B T_c$ is of the order of the exchange self energy
$0.1$eV.  However, quantum fluctuations may reduce this value. A
spin-polarized CDW and a Luttinger liquid are possible true
groundstates. It may be  worthwhile to investigate these issues
using improved approximation schemes than EMA employed in this
paper, such as spin density functional method and exact
diagonalization techniques.  Our results suggest that experimental
investigations of magnetism in one dimensional electron gas of an
armchair ribbon may produce numerous interesting results.

\begin{acknowledgements}
We thank A.H. MacDonald for several useful suggestions.  This
research was supported by Basic Science Research Program through the
National Research Foundation of Korea(NRF) funded by the Ministry of
Education, Science and Technology (2012R1A1A2001554).
\end{acknowledgements}

\end{document}